\documentclass[aps,prl,reprint,superscriptaddress,showpacs]{revtex4-1}

\usepackage{graphicx}
\usepackage{amssymb,amsmath}
\usepackage{bm}
\usepackage{dcolumn}

\begin{document}

\title{Multiscale quantum-defect theory and its application to atomic spectrum}
\author{Haixiang Fu}
\affiliation{College of Information Science and Engineering, Huaqiao University,
Xiamen, Fujian 361021, China}
\author{Mingzhe Li}
\affiliation{Department of Physics and Institute of Theoretical Physics and Astrophysics, %
Xiamen University, Xiamen, Fujian 361005, China}
\author{Meng Khoon Tey}
\author{Li You}
\affiliation{Collaborative Innovation Center of Quantum Matter, Beijing, China}
\affiliation{State Key Laboratory of Low-Dimensional Quantum Physics, Department of
Physics, Tsinghua University, Beijing 100084, China}
\author{Bo Gao}
\email{bo.gao@utoledo.edu}
\affiliation{Collaborative Innovation Center of Quantum Matter, Beijing, China}
\affiliation{Department of Physics and Astronomy, Mailstop 111, University of Toledo,
Toledo, Ohio 43606, USA}
\date{\today}

\begin{abstract}
We present a multiscale quantum-defect theory based on the first analytic solution for a two-scale long range potential consisting of a Coulomb potential and a polarization potential. In its application to atomic structure, the theory extends the systematic understanding of atomic Rydberg states, as afforded by the standard single-scale quantum-defect theory, to a much greater range of energies to include the first few excited states and even the ground state. Such a level of understanding has important implications not only on atomic structure, but also on the electronic structure of molecules and on atomic and molecular interactions and reactions. We demonstrate the theory by showing that it provides an analytic description of the energy variations of the standard Coulomb quantum defects for alkali-metal atoms.
\end{abstract}

\maketitle




The Rydberg-Ritz formula \cite{Ritz1903,Hartree1928} for atomic spectrum 
\begin{equation}
(E_{nlj}-E_\mathrm{ion})/R_M = -\frac{1}{(n-\mu^{\text{Coul}}_{lj})^2} \;,
\label{eq:Ryd}
\end{equation}
was one of the very first universal properties uncovered for quantum systems, and played an important role in the very establishment of the quantum theory. Here $E_\mathrm{ion}$ represents the ionization energy, $R_M=R_\infty/(1+m_e/M)$ is the reduced mass Rydberg constant (with $M$ being the ion mass), $n$ is the principle quantum number, $l$ and $j$ are the orbital and total angular momenta of the electron, respectively, and $\mu^{\text{Coul}}_{lj}$ is the Coulomb quantum defect. The formula asserts that despite considerable differences in atomic spectra, the Rydberg series for different atoms differ from each other only in a quantum defect which encapsulates all complexities of short-range interactions. This universality, which originates from the fact that a sufficiently highly excited electron sees mostly the Coulomb potential, stimulated the development of the quantum-defect theory (QDT) and multichannel quantum-defect theory (MQDT) \cite{sea83,Greene1985,aym96,Jungen1996}. They have long become the standard for understanding atomic and molecular spectra and electron-ion interactions \cite{sea83,Greene1985,aym96,Jungen1996,Burke2011}.

The universality as represented by the Rydberg-Ritz formula with a constant $\mu^{\text{Coul}}_{lj}$ is however strictly applicable only to sufficiently highly excited Rydberg states. This is reflected, especially for atoms with highly polarizable cores, by a significant energy dependence of $\mu^{\text{Coul}}_{lj}$ for lower lying states (see, e.g., Refs.~\cite{Lorenzen1983,Weber1987,Sanguinetti2009,Mack2011}). Does there exist a more general  universality that applies also to the first few excited states or even the ground state, which are often of more practical and experimental interest? The answer to this question has implications far beyond atomic structure. Not only will it determine the degree to which other single-atom properties, such as oscillator strengths, ionization cross sections, static and dynamic polarizabilities (see, e.g., Ref.~\cite{mit10}), follow universal behaviors outside of the Rydberg regime. It will also determine the degree to which important atomic interaction parameters, such as the $C_6$ van der Waals coefficients (see, e.g., Ref.~\cite{Derevianko1999}), follow universal behaviors for different atoms and different electronic states. Even further, it will determine the relations among different electronic states of a molecule and the relations among electronic states of different molecules. The prospect for such systematic understanding of an entire manifold of molecular electronic states can be crucial for understanding atomic and molecular interactions and reactions, especially in excited electronic states where many of them participate simultaneously.

This work establishes a broader universality in atomic structure as the first application of an analytic \textit{two-scale} QDT. The theory is based on our analytic solution for the ``Coulomb+Polarization'' potential of the form $-C_1/r-C_4/r^4$. It is, to the best of our knowledge, the first analytic solution of the Sch\"{o}dinger equation for a two-scale central potential, for which no analytic solution is previously known or expected. Our solution represents a new class of special functions and has potential for generalization to other multiscale potentials. In applying the two-scale QDT to atomic structure, we show that the theory introduces a ``new'' quantum defect that has much weaker energy dependence than the traditional Coulomb quantum defects $\mu^{\mathrm{Coul}}$ that goes into the Rydberg-Ritz formula, and the theory provides an analytic description of the energy dependence of $\mu^{\mathrm{Coul}}$ for Rb and Cs atoms, down to their ground states. The results establishes, in an analytic framework, a broader universality in atomic structure and spectrum. Physically, it means that the statement that a Rydberg electron sees mostly the Coulomb potential can be replaced by a more general statement that an excited or an outer electron sees mostly a Coulomb potential plus a polarization potential, to a remarkable accuracy. 

We note that the importance of polarization in atomic structure is well known, as reflected both in perturbation calculations for alkali metals \cite{Freeman1976,Drake1991}, and in model potentials chosen, e.g., both for ``one-electron'' alkali atoms \cite{Weisheit1972,Norcross1973} and for ``two-electron'' alkaline earth atoms \cite{Greene1987,aym96}. The underlying universality was however difficult to identify, define, or describe, since a perturbative treatment is only applicable to high angular momentum states, and the effect of core polarization is difficult to distinguish from other short-range effects in a numerical calculation.


Our two-scale QDT for ``Coulomb+Polarization'' potential is built upon the analytic solution of the radial Schr\"{o}dinger equation
\begin{equation}
\left[-\frac{\hbar^2}{2\mu}\frac{d^2}{dr^2}+\frac{\hbar^2l(l+1)}{2\mu r^2} -%
\frac{C_1}{r}-\frac{C_4}{r^4}-\epsilon\right] v_{\epsilon l}(r) = 0 \;,
\label{eq:sch1a4a}
\end{equation}
where $\mu$ is the reduced mass. It has two characteristic length scales $\beta_n := (2\mu C_n/\hbar^2)^{1/(n-2)}$ ($n=1,4$), corresponding to each one of the potential terms of the form of $-C_n/r^n$ ($n=1,4$). Each length scale $\beta_n$ has a corresponding energy scale of $s^{(n)}_E = (\hbar^2/2\mu)(1/\beta_n)^2$. Scaling the radius $r$ by $\beta_1$, and the energy $\epsilon$ by its corresponding $s_E^{(1)}$, the scaled Schr\"{o}dinger equation takes the form of  
\begin{equation}
\left[\frac{d^2}{dr_s^2} - \frac{l(l+1)}{r_s^2} + \frac{1}{r_s} +
(\beta_4/\beta_1)^2\frac{1}{r_s^4} + \epsilon_s\right] v_{\epsilon_s l}(r_s)
= 0 \;, 
\label{eq:ssch1a4a}
\end{equation}
where $r_s := r/\beta_1$ and $\epsilon_s := \epsilon/s_E^{(1)}$, with the ratio of length scales $\beta_4/\beta_1$ serving as a measure of the relative strength of the Coulomb and the polarization potentials.

Through generalizations of techniques that have led to other single-scale QDT solutions for $1/r^6$ \cite{gao98a}, $1/r^3$ \cite{gao99a}, and $1/r^4$ \cite{hol73,wat80,gao10a,idz11,Gao2013c} potentials, we have solved Eq.~(\ref{eq:ssch1a4a}) analytically to obtain its QDT base pair of solutions and the corresponding QDT functions \cite{gao08a}. The base pair $f^c$ and $g^c$ are defined with energy and partial wave independent asymptotic behaviors around the origin (specifically $r_s\ll \beta_4/\beta_1)$,
\begin{eqnarray}
f^c_{\epsilon_s l}(r_s) &\sim& r_s\sqrt{\frac{2}{%
\pi(\beta_4/\beta_1)}} \cos\left[(\beta_4/\beta_1)/r_s-\pi/4 \right] \;,
\label{eq:fcdef4} \\
g^c_{\epsilon_s l}(r_s) &\sim& -r_s\sqrt{\frac{2}{%
\pi(\beta_4/\beta_1)}} \sin\left[(\beta_4/\beta_1)/r_s -\pi/4 \right] \;,
\label{eq:gcdef4}
\end{eqnarray}
for all energies $\epsilon_s$ \cite{gao01,gao08a}. They are normalized such that their Wronskian $W(f^c,g^c) := f^c(d g^c/dr_s)-(d f^c/dr_s)g^c = 2/\pi$. For $\epsilon_s<0$, we have shown that the QDT base pair has asymptotic behavior at large $r_s$ given by 
\begin{align}
f^c_{\epsilon_s l}(r_s) \overset{r_s\rightarrow \infty}{\sim}& \frac{1}{%
\sqrt{\pi\kappa_s}}\left[ W^c_{f+}(2\kappa_s
r_s)^{1/(2\kappa_s)}e^{-\kappa_s r_s}\right.  \notag \\
&+\left.W^c_{f-}(2\kappa_s r_s)^{-1/(2\kappa_s)}e^{+\kappa_s r_s}\right] \;,
\label{eq:fclrne} \\
g^c_{\epsilon_s l}(r_s) \overset{r_s\rightarrow \infty}{\sim}& \frac{1}{%
\sqrt{\pi\kappa_s}}\left[ W^c_{g+}(2\kappa_s
r_s)^{1/(2\kappa_s)}e^{-\kappa_s r_s} \right.  \notag \\
&+\left.W^c_{g-}(2\kappa_s r_s)^{-1/(2\kappa_s)}e^{+\kappa_s r_s}\right] \;,
\label{eq:gclrne}
\end{align}
where $\kappa_s=(-\epsilon_s)^{1/2}$. It gives a $2\times 2$ $W^c$ matrix with elements $W^c_{xy}$ describing the evolution of a wave function through the  $-C_1/r-C_4/r^4$ potential at negative energies \cite{gao08a}.


In terms of the $W^c$ elements, the bound spectrum for any potential $V(r)$ that behaves asymptotically as  $-C_1/r-C_4/r^4$ can be formulated \cite{gao08a} as the solutions of
\begin{equation}
\chi^{c}_l(\epsilon_s,\beta_4/\beta_1) = K^c(\epsilon,l,j) \;.
\label{eq:qdtbsp}
\end{equation}
Here $\chi^{c}_l = W^c_{f-}/W^c_{g-}$ is a universal function of the scaled energy $\epsilon_s$ and the ratio of length scales $\beta_4/\beta_1$. The $K^c$ is a short-range $K$ matrix defined by matching the short-range wave function, $u_{\epsilon l}(r)$ for potential $V(r)$, to a linear combination of the QDT base pair \cite{gao01,gao08a}
\begin{equation}
u_{\epsilon l j}(r) = A_{\epsilon l j}[f^c_{\epsilon_s l}(r_s) - K^c(\epsilon,l,j)
g^c_{\epsilon_s l}(r_s)]\;,  \label{eq:wfn}
\end{equation}
at any radius where $V(r)$ has become well represented by $-C_1/r-C_4/r^4$. Compared to single-scale QDT formulations \cite{gao08a}, the formula for bound spectrum in the two-scale formulation, Eq.~(\ref{eq:qdtbsp}), is structurally the same except that the two-scale $\chi^{c}_l$ depends parametrically on $\beta_4/\beta_1$, and the two-scale $K^c$ is defined in reference to $-C_1/r-C_4/r^4$, instead of $-C_1/r$ solutions. 

From our analytic solutions, we obtain
\begin{equation}
\chi^{c}_l = \frac{\tan\theta^-_l+ \tan(\pi\nu/2)(1+\widetilde{M}_{l})/(1-%
\widetilde{M}_{l})} {1-\tan\theta_l^- \tan(\pi\nu/2)(1+\widetilde{M}_{l})/(1-%
\widetilde{M}_{l})} ,  \label{eq:chic}
\end{equation}
Here $\nu$ is the characteristic exponent for the $-1/r_s-(\beta_4/\beta_1)^2/r_s^4$ potential, given in the supplemental material \cite{sup}. It is a function of the scaled energy $\epsilon_s$ and depends parametrically on $\beta_4/\beta_1$, as are the $\tan\theta^-_l$ and $\widetilde{M}_{l}$ functions in Eq.~(\ref{eq:chic}). The function $\tan\theta^-_l$ is given by $ \tan\theta^-_l=Y^-_{l}/X^-_{l}$, with $X^-_l =\sum_{j=-\infty}^\infty (-1)^j b^-_{2j} $, and $Y^-_l =\sum_{j=-\infty}^\infty (-1)^j b^-_{2j+1} $. Here $b^-_m$s are the coefficients for the generalized Neumann expansion \cite{Cavagnero1994,gao98a} of a wave function. They are given by $b^-_0=1$, and
\begin{multline}
b^-_m = \Delta^m\frac{\Gamma(\nu)\Gamma(\nu+\nu_0+1)\Gamma(\nu-\nu_0+1)}
	{\Gamma(\nu-w^-)} \\
	\times\frac{\Gamma(\nu+m-w^-)}{\Gamma(\nu+m)\Gamma(\nu+\nu_0+m+1)\Gamma(\nu-\nu_0+m+1)}
	c_m(\nu) \;, 
\end{multline}
\begin{multline}
b^-_{-m} = (-1)^m\Delta^m\frac{\Gamma(\nu+1+w^-)}
	{\Gamma(\nu+1)\Gamma(\nu+\nu_0)\Gamma(\nu-\nu_0)} \\
	\times\frac{\Gamma(\nu-m+1)\Gamma(\nu+\nu_0-m)\Gamma(\nu-\nu_0-m)}{\Gamma(\nu+1+w^--m)}
	c_m(-\nu) \;,
\end{multline}
in which $m$ is a positive integer, $\nu_0:=l+1/2$, $\Delta:=(\beta_4/\beta_1)\sqrt{-\epsilon_s}$, $w^-:=-1/2+1/(2\sqrt{-\epsilon_s})$, and $c_m(\nu) := \prod_{j=0}^{m-1}Q(\nu+j)$ with $Q(\nu)$ given by a continued fraction   
\begin{equation}
Q(\nu) = \frac{1}{1-(\frac{\beta_4}{\beta_1})^2\frac{(\nu+3/2)^2\epsilon_s+1/4} {%
(\nu+1)[(\nu+1)^2-\nu_0^2](\nu+2)[(\nu+2)^2-\nu_0^2]}Q(\nu+1)} \;.
\label{eq:Qcf}
\end{equation}
The $\widetilde{M}_l$ function in Eq.~(\ref{eq:chic}) is given by $\widetilde{M}_l := [\Gamma(\nu-w^-)/\Gamma(-\nu-w^-)]M_l $, where 
\begin{multline}
M_{l} := \left[(\beta_4/\beta_1)^2|\epsilon_s|\right]^{\nu} \\
\times\frac{\Gamma(1-\nu)\Gamma(1-\nu+\nu_0)\Gamma(1-\nu-\nu_0)} {%
\Gamma(1+\nu)\Gamma(1+\nu+\nu_0)\Gamma(1+\nu-\nu_0)} \left(\frac{C_{l}(-\nu)%
}{C_{l}(+\nu)}\right) \;,  
\label{eq:Ml}
\end{multline}
in which $C_{l}(\nu) := \prod_{j=0}^{\infty}Q(\nu+j)$.

\begin{figure}
\includegraphics[width=\columnwidth]{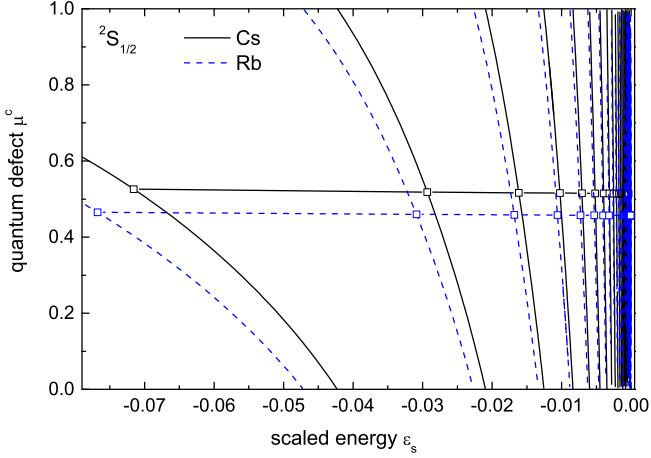} 
\caption{(Color online). The $\mu^c$ representation of the multiscale QDT bound spectrum, as the crossing points between $\chi^{\mu}_l(\epsilon_s,\beta_4/\beta_1)$ (the curves plotted) and the function $\mu^c(\epsilon,l,j)$ (the nearly horizontal lines), illustrated here for $^2S_{1/2}$ ($l=0$) series of Cs (Black) and Rb (Blue). 
When the spectrum is known, the function $\chi^{\mu}_l$ evaluated at bound state energies gives a discrete representation of the $\mu^c$ function. The experimental data for Cs \cite{Weber1987} and Rb \cite{Sansonetti2006,Mack2011} (squares) show that $\mu^c$, and therefore $K^c$, are to a very good approximation a constant, including those for the ground states ($6s$ for Cs and $5s$ for Rb). The $\chi^\mu$ as a function energy is different for Cs and Rb due to their different length scale ratios $\beta_4/\beta_1$, which in turn is due  primarily to their different core polarizabilities. $\alpha_\mathrm{core}$ is taken here to be 15.77 a.u. \cite{Weber1987} for Cs, and 9.076 a.u.  for Rb \cite{Drake1991}.
\label{fig:chimu}}
\end{figure}

Equation~(\ref{eq:qdtbsp}) gives the two-scale QDT spectrum as the cross points between a universal function $\chi^c_l(\epsilon_s,\beta_4/\beta_1)$ and a short-range $K^c(\epsilon,l,j)$ function. It can also be formulated in a $\mu^c$ representation \cite{Gao2013c}, as $\chi^{\mu}_l(\epsilon_s,\beta_4/\beta_1) = \mu^c(\epsilon,l,j)$, where $\chi^\mu_l :=[\tan^{-1}(\chi^c_l)-\pi/4]/\pi$ and $\mu^c$ is the ``new'' quantum defect defined by $\mu^c :=[\tan^{-1}(K^c)-\pi/4]/\pi$. Both $\chi^\mu_l$ and $\mu^c$ are taken to be within a range of $[0,1)$ by taking $\tan^{-1}(x)$ to be within a range of $[\pi/4, 5\pi/4)$. In applying Eq.~(\ref{eq:qdtbsp}) to the spectra of an alkali atom, one has $C_1=1$ and $C_4 = \alpha_{\mathrm{core}}/2$ in atomic units, where $\alpha_{\mathrm{core}}$ is the polarizability of the core (i.e. the ionic core excluding the outer electron), and $\beta_4/\beta_1=2\mu^{3/2}\sqrt{\alpha_{\mathrm{core}}}$. 

The key difference between the two-scale QDT and the standard single-scale QDT \cite{sea83,aym96} lies in the fact that the ``new'' quantum defect $\mu^c$, being defined in reference to the $-C_1/r-C_4/r^4$ potential, is determined by the logarithmic derivative of the wave function at much shorter distances than the standard quantum defect $\mu^{\text{Coul}}_{lj}$ defined in reference to the $-C_1/r$ Coulomb potential, and has therefore much weaker energy dependence. Figure~\ref{fig:chimu} illustrates both the $\mu^c$ representation \cite{Gao2013c} of the two-scale QDT spectrum, and the weak energy dependence of $\mu^c(\epsilon,l,j)$, using experimental data for the $^2S_{1/2}$ series of both Cs \cite{Weber1987} and Rb \cite{Sansonetti2006,Mack2011}.

From a different perspective, the two-scale QDT provides an analytic description of the energy dependence of the $\mu^{\text{Coul}}_{lj}$  in the Rydberg-Ritz formula. Specifically, Eq.~(\ref{eq:qdtbsp}) for the spectrum can be solved, more precisely re-casted, as the solutions of 
\begin{equation}
\epsilon_{nljs} = -\frac{1}{4 [n-\mu^{\text{Coul}}_{lj}(\epsilon_{nljs})]^2} \;.
\label{eq:Ryds}
\end{equation}
Here $\epsilon_{nljs}$ is a scaled bound state energy (hence the subscript $s$) defined by $\epsilon_{nljs}:=(E_{nlj}-E_\mathrm{ion})/s_E^{(1)}=(E_{nlj}-E_\mathrm{ion})/4R_M$,
\begin{equation}
\tan(\pi\mu^{\text{Coul}}_{lj}) = -\frac{1-t_l}{1+t_l}\tan[\pi(\nu-1/2)] \;,
\label{eq:cqd}
\end{equation}
with 
\begin{multline}
t_l = \left[\frac{\Gamma(1/(2\kappa_s)+\nu+1/2)}{\Gamma(1/(2\kappa_s)-%
\nu+1/2)}M_l\right] \\
\times\frac{K^c\cos(\pi\nu/2-\theta^-_l)+\sin(\pi\nu/2-\theta^-_l)}
{K^c\cos(\pi\nu/2+\theta^-_l)-\sin(\pi\nu/2+\theta^-_l)} \;.
\end{multline}
Equation~(\ref{eq:Ryds}), which is formally equivalent to Eq.~(\ref{eq:Ryd}) with a different scaling, shows that the spectrum for any potential that behaves asymptotically as $-C_1/r-C_4/r^4$ can be expressed as a Rydberg-Ritz formula with an energy-dependent $\mu^{\text{Coul}}_{lj}$, consistent with the general conclusion by Hartree many years ago \cite{Hartree1928}. The energy dependence is described analytically by Eq.~(\ref{eq:cqd}). It is, to the best of our knowledge, the first nonperturbative analytic description of this energy dependence. 

\begin{figure}[tpbh]
\includegraphics[width=1.1\columnwidth]{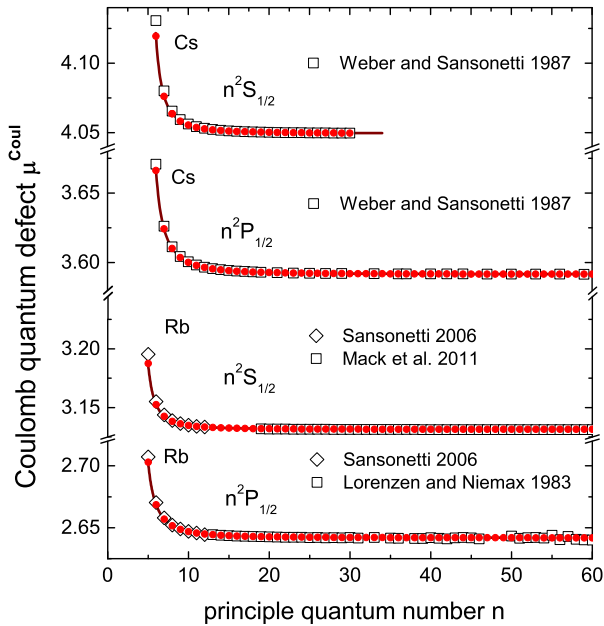} 
\caption{(Color online). Energy variation of the Coulomb quantum defects starting from the ground states. Open symbols: data derived from experimental spectra \cite{Lorenzen1983,Weber1987,Sansonetti2006,Mack2011} and Eq.~(\ref{eq:Ryds}). Solid dots: QDT predictions from Eq.~(\ref{eq:cqd}) with constant $K^{c}$s, Connecting lines: Eq.~(\ref{eq:cqd}) which is well defined for all negative energies.
\label{fig:cqdSP}}
\end{figure}

Figure \ref{fig:cqdSP} compares the Coulomb quantum defects $\mu^{\text{Coul}}_{lj}$ to the two-scale QDT predictions of Eq.~(\ref{eq:cqd}) with a \textit{constant} $K^c$, using as examples the $^2S_{1/2}$ and $^2P_{1/2}$ series of Rb and Cs. The Rb data are taken from recent precision measurement and analysis in Refs.~\cite{Lorenzen1983,Sansonetti2006,Mack2011}. The $\alpha_\mathrm{core}$ is taken to be 9.076 a.u. ~\cite{Drake1991}. The dimensionless parameter $K^c$ is determined to be $-1.314$ for the $^2S_{1/2}$ series, and $-1.173$ for the $^2P_{1/2}$ series, by fitting to an intermediate-$n$ portion of the spectra. They correspond to $\mu^c=0.4571$ for the $^2S_{1/2}$ series, and $\mu^c=0.4747$ for the $^2P_{1/2}$ series.
The Cs spectrum data are from Ref.~\cite{Weber1987}, and we have taken its $\alpha_{\mathrm{core}}= 15.77$ a.u. ~\cite{Weber1987}. The parameters $K^{c}$ are determined to be $-0.9135$ for the $^2S_{1/2}$ series, and $-0.8942$ for the $^2P_{1/2}$ series. They correspond to $\mu^c=0.5144$ and $\mu^c=0.5178$ for the $^2S_{1/2}$ and $^2P_{1/2}$ series, respectively. It is remarkable that even with such constant $K^c$s (or $\mu^c$s), the two-scale QDT predicts $\mu^{\text{Coul}}_{lj}$ with an accuracy better than 0.3\% for the ground states, and the energy with an accuracy better than 1.3\% for the ground states, and progressively better for excited states.

The weak energy dependence of $K^c$ (or $\mu^c$), and the degree to which a constant $K^c$ (or $\mu^c$) describes the energy dependence of $\mu^{\text{Coul}}_{lj}$ show that atomic spectrum follows the universal behavior as characterized by the $-C_1/r-C_4/r^4$ solution, not only for high partial wave states \cite{Freeman1976,Drake1991}, but also for the $S$ and $P$ states, and not only for highly excited states, but also for the first few excited states and the ground state. The weak energy dependence of $K^c$ also implies that the probability for finding this outer electron in the region where the potential defers substantially from $-C_1/r-C_4/r^4$ is small \cite{gaoupb}, and the wave function, including its normalization, is accurately given by the analytic $-C_1/r-C_4/r^4$ wave function. This combination of spectrum and wave function both following a broader universal behavior is what will lead to the broader universal behaviors in atomic polarizability and the $C_6$ coefficient for different atoms and different electronic states.

Through the weak energy dependence of its short-range parameters, the two-scale QDT allows the determination of the Rydberg spectrum from the measurement of the first few excited states, and allows the spectral determination of the core polarizability \cite{gao01}. Above the ionization threshold, the same theory provides an analytic description of electron-ion scattering \cite{sea83,aym96,Burke2011} over a wide range of energies. Higher accuracy on the spectrum and other atomic properties, when desired, can be achieved by taking into account the weak energy dependence $K^c$ (or $\mu^c$) using a standard Taylor expansion (since they are analytic functions of energy, unlike $\mu^{\mathrm{Coul}}$ in the presence of the polarization potential) \footnote{Relativistic effects are mostly short-range (where electron is moving faster) effects that are reflected, e.g., in different $\mu^c$ for different $j$ states. In our formulation, they can be determined from experimental spectra without explicitly solving the relativistic equations \cite{Johnson2008,Safronova2011}.}. Multichannel \cite{sea83,aym96,gao05a} and anisotropic generalizations of the theory will extend its description to atomic species other than group-I atoms. The two-scale QDT can also be used in a fully \textit{ab initio} fashion together with a $R$-matrix theory \cite{aym96,Burke2011}, leading to more efficient and accurate calculations with a smaller $R$-matrix box.


In conclusion, we have presented a two-scale QDT for a Coulomb plus polarization potential, and have used it to establish a broader universality in atomic spectrum, covering not only the Rydberg states, but also lower lying states including the ground state. 
Beyond single-atom and two-atom properties (such as $C_6$), the theory will have implications on molecular electronic structure, and on atomic and molecular interactions and reactions. For molecular structure, for instance, it implies that the electronic structures, including both ground and excited bands, of Cs$_N$ and Rb$_N$ clusters differ from each other only through different polarizabilities of Cs$^+$ and Rb$^+$ ions and a few quantum defects $\mu^c(l,j)$ for Cs and Rb atoms (with $N\rightarrow\infty$ corresponding to condensed phases). Mathematically, the same $-C_1/r-C_4/r^4$ solution is applicable not only to electron-ion, but also to ion-ion interactions. Finally, this first establishment of analytic multiscale QDT gives hope that similar theories can be developed for other interactions, such as atom-atom \cite{chi10} and ion-atom \cite{gao10a,idz11,Gao2013c,LYG2014}, that are currently treated either at only a single scale \cite{gao05a,gao08a}, or at multiscale but only numerically (see, e.g., Refs.~\cite{mie84a,tie93,Burke1998}). If successful, such developments will have impact on almost every aspect of atomic and molecular structure, interactions, and reactions.

\begin{acknowledgments}
We thank Prof. Lee Zhang for helpful discussions. This work is supported by MOST 2013CB922004 of the National Key Basic Research Program of China, by  NSFC (No.~91121005, No.~10947010, No.~11374176, and No.~11328404), and by HuaQiao University Fund (No.~09Y0155). The work at Toledo is supported by NSF (PHY-1306407).
\end{acknowledgments}

\bibliography{bgao,qdt,atom,atomAtom,Rydberg,ionAtom,Fesh}

\end{document}